\documentclass[sigconf,nonacm]{acmart}

\renewcommand\footnotetextcopyrightpermission[1]{} 

\AtBeginDocument{%
  }

\copyrightyear{2025}
\acmYear{2025}
\setcopyright{acmlicensed}
\acmConference[MMFood '25]{Proceedings of the 1st International Workshop on Multi-modal Food Computing}{October 27--28, 2025}{Dublin, Ireland}
\acmBooktitle{Proceedings of the 1st International Workshop on Multi-modal Food Computing (MMFood '25), October 27--28, 2025, Dublin, Ireland}
\acmDOI{10.1145/3746264.3760498}
\acmISBN{979-8-4007-2046-8/2025/10}

\acmISBN{978-1-4503-XXXX-X/2018/06}

\begin{document}

\title{What’s Not on the Plate? Rethinking Food Computing through Indigenous Indian Datasets}

\author{Pamir Gogoi}
\authornote{Corresponding Authors}
\email{pamir.gogoi@karya.in}
\orcid{0000-0002-4077-8488}
\author{Neha Joshi}
\authornotemark[1]
\email{neha@karya.in}
\orcid{0009-0000-8785-9296}
\author{Ayushi Pandey}
\authornotemark[1]
\email{ayushi@karya.in}
\orcid{0009-0006-8307-281X}
\author{Vivek Seshadri}
\orcid{0009-0000-4388-4246}
\affiliation{%
  \institution{Karya Inc.}
  \city{Bengaluru}
  \state{Karnataka}
  \country{India}
}

\author{Deepthi Sudharsan}
\orcid{0000-0002-1990-3010}
\author{Kalika Bali}
\orcid{0000-0001-9275-742X}

\affiliation{%
  \institution{Microsoft Research}
  \city{Bengaluru}
  \state{Karnataka}
  \country{India}}

\author{Saransh Kumar Gupta}
\email{saransh.gupta@ashoka.edu.in}
\orcid{0009-0000-5887-2301}
\authornotemark[1]

\author{Lipika Dey}
\orcid{0000-0003-3831-5545}

\author{Partha Pratim Das}
\orcid{0000-0003-1435-6051}

\affiliation{%
  \institution{Ashoka University}
  \city{Sonepat}
  \state{Haryana}
  \country{India}
}







\renewcommand{\shortauthors}{Gogoi et al.}

\begin{abstract}
This paper presents a multimodal dataset of 1,000 indigenous recipes from remote regions of India, collected through a participatory model involving first-time digital workers from rural areas. The project covers ten endangered language communities in six states. Documented using a dedicated 
mobile app, the data set includes text, images, and audio, capturing traditional food practices along with their ecological and cultural contexts. This initiative addresses gaps in food computing, such as the lack of culturally inclusive, multimodal, and community-authored data. By documenting food as it is practiced rather than prescribed, this work advances inclusive, ethical, and scalable approaches to AI-driven food systems and opens new directions in cultural AI, public health, and sustainable agriculture.
\end{abstract}





\maketitle

\section{Introduction}
Food computing encompasses a wide range of computational techniques designed to address food-related challenges across domains such as medicine \cite{boanos2017investigation}, nutrition \cite{jiang2024food}, safety \cite{jin2020big}, and agriculture \cite{castello2018personal}. Common research directions include the processing of massive volumes of heterogeneous data, ranging from recipe books and food images to cooking videos and food blogs. Advances in computer vision have enabled automatic recognition of dishes, ingredients, and even cooking stages from visual inputs, which support dietary monitoring, culinary education, and food documentation \cite{gomes2012applications, kakani2020critical, rokhva2024computer}. Similarly, natural language processing techniques have significantly improved recipe understanding, ingredient substitution, and cross-cultural food translation, powering applications in personalized nutrition, food recommendation systems, and culinary knowledge graphs \cite{fried2014analyzing, rezayi2024chatdiet, zhou2024foodsky}. 

Food in India is deeply embedded in everyday life—from rites to medicine, and agriculture to socialization. Yet, despite this vast and living knowledge system, there are only a few notable efforts within the Indian context in the field of food computing. Research initiatives like IIIT Delhi’s Computational Gastronomy Lab \cite{bagler_iiitd_computationalgastronomy} and Ashoka University’s Food Computing Lab \cite{foodcomputing_ashoka} are pioneering but early steps toward modeling the unique cultural, linguistic, and culinary landscape of Indian food traditions. Existing methods, datasets, and Internet-based data collection fail to capture the socio-ecological and cultural richness embedded in Indian home cooking. With over 730 registered tribal groups, India offers a cultural context that is dramatically underrepresented in current computational resources. As a result, food computing suffers from a severe representational bias. 

We describe the documentation of 1,000 multimodal recipes from remote, indigenous regions in India. These recipes are captured by first-time digital workers who serve as critical links between ancestral knowledge, their digitalization, and technology platforms. The resulting dataset includes video, audio, images, text, and translations, capturing the representation of Indian culinary practice, suitable for alignment with the ontology and extension of FKG.in—India’s first comprehensive food knowledge graph that aims to integrate AI, domain-specific ontologies, and real-world cooking intelligence with health, nutritional, and agroecological information and contexts. While FKG.in is being developed independently as an unrelated initiative, this paper brings together contributors from both efforts to explore their convergence among other potential directions.

This paper explains the key aspects of data collection and use this dataset to highlight major gaps related to food computing in India. We present this dataset as a starting point for identifying key challenges, exploring potential applications, and underscoring the need for scalable, inclusive, and nuanced food datasets rooted in the Indian context. 

\section{Related Work and Gaps}
\label{sec:related_work}

Food computing research has produced several large-scale datasets and models focused on visual recognition, cross-modal retrieval, and recipe understanding \cite{min2019survey}. Benchmark multimodal datasets such as Recipe1M+ \cite{marin2019learning}, Food-101 \cite{bossard14}, and VireoFood-172 \cite{VireoFood172} have advanced tasks like classification, image-to-recipe generation, and ingredient prediction, but rely mainly on standardized, internet-based content. Most efforts in Europe, China, and the US focus on industrially structured recipes, processed food labels, or scientific nutrition databases, resulting in cultural bias, limited ecological diversity, and underrepresentation of vernacular practices, especially in low-resource settings \cite{MIN2022100484, thames2021nutrition5k, obrist2018future}. 
Recent work on culinary knowledge graphs \cite{haussmann2019foodkg}, dietary monitoring \cite{myers2015Im2calories}, fine-grained datasets \cite{li2024foodieqa, magomere2025world}, and culturally aware modeling \cite{zhou2024does, cao2024cultural, foods9060823} uses readily available data, but oral and hyperlocal food knowledge—especially in India—remains largely absent from AI.

India-specific datasets include INDoRI \cite{khanna2023indoriindiandatasetrecipes} (\textasciitilde5,200 recipes, 18 regions), image-based resources like the Indian Spices Image Dataset \cite{THITE2024110936} (\textasciitilde11,000 images, 19 spices), IndianFoodNet \cite{agarwal2023indianfoodnet} (\textasciitilde5,500 images, 30 classes), and IndianFood10/20 \cite{pandey2022objectdetectionindianfood}, as well as structured compilations like RecipeDB \cite{batra2020recipedb} and FKG.in \cite{Gupta2024FKG}. While valuable, these are mostly single-modal, crowd- or web-sourced, and lack cultural or ecological grounding. CookingINWild \cite{khanna2024cookinginwild} adds multimodal video of Indian cooking, but remains limited to visual data without deeper community-anchored or narrated content.

Beyond the core domain of food computing, adjacent fields such as agroecology, ethnobotany, and digital humanities have long recognized the value of participatory knowledge production, especially in the context of local food systems. 
Ethnographic research, such as work on Himalayan indigenous communities \cite{das2022indigenous}, documents how local food practices intertwine with ecology and climate change. Similarly, multimedia archives of oral recipes and seasonal foraging patterns exist in regional ethnobotany \cite{banisetti2023digital, mir2024kashmir}, but are rarely digitized or integrated into AI frameworks. These lay important precedents for community-driven, multimodal documentation, but they have yet to inform food computing datasets.

An overview of the existing work and remaining gaps in the food knowledge documentation literature across selected dimensions is detailed in Table \ref{tab:gap_comparison}. Our data collection process and the dataset tries to address these gaps by creating a sensitive and ethical space for documenting indigenous knowledge, culinary practices, and local seasonal ingredients. We try to between globally commodified food data and real-world, culturally rooted food knowledge by building a multimodal, locally grounded, and community-led dataset that not only fills data gaps but also opens new directions for equitable, scalable, and culturally nuanced AI systems. 

\begin{table*}
  \caption{Overview of Existing Work and Remaining Gaps in Food Knowledge Documentation}
  \label{tab:gap_comparison}
  \begin{tabular}{p{0.1cm}p{3.3cm}p{3.8cm}p{9.2cm}}
    \toprule
    \multicolumn{1}{c}{\textbf{No.}} & \multicolumn{1}{c}{\textbf{Dimension}} & \multicolumn{1}{c}{\textbf{Existing Work}} & \multicolumn{1}{c}{\textbf{Remaining Gaps}} \\
    \midrule
    1 & {\centering Modalities} 
    & {\centering Images, text, video action clips} 
    & {\centering Missing audio narration, step-by-step process video, field metadata} \\
    2 & {\centering Cultural Grounding} 
    & {\centering Web/crowdsourced recipes} 
    & {\centering No indigenous or traditional area-based documentation} \\
    3 & {\centering Scalability} 
    & {\centering Large, shallow datasets} 
    & {\centering Need for deep, qualitative data from sampled communities} \\
    4 & {\centering Community Participation} 
    & {\centering Generic crowdsourced inputs} 
    & {\centering Missing specific, local, and lived food knowledge} \\
    5 & {\centering Knowledge Integration} 
    & {\centering Limited use of ontologies/KGs} 
    & {\centering Direct linkage/extensibility to KGs and specialized AI reasoning systems} \\
    6 & {\centering Contextual Factors} 
    & {\centering Formulaic and static recipes} 
    & {\centering Missing temporal, ecological, and oral transmissions of food practices} \\
    7 & {\centering Language \& Access} 
    & {\centering English-dominant data} 
    & {\centering Recipes in regional languages, with translations and transliteration} \\
    8 & {\centering Ethical Data Practices} 
    & {\centering Extractive, unclear consent} 
    & {\centering Participatory design with attribution, consent, and fair labor models} \\
    \bottomrule
  \end{tabular}
\end{table*}

\section{A Multimodal Recipe Dataset from Indigenous Indian Regions }
\label{sec:multimodal_dataset}



\subsection{The Recipe Dataset Project}
\label{subsec:recipe_project}

We adopt a ground-up digital methodology, where multimodal data is collected from Indigenous communities. This expands the scope of computational food data to include locally produced, culturally situated knowledge, positioning community members as active knowledge producers rather than passive subjects. The resulting dataset is culturally grounded and computationally usable, bridging participatory knowledge systems with machine-readable infrastructures. 

\subsubsection{Project Overview - }
\label{subsubsec:project_overview}
This initiative successfully collected approximately 1,000 traditional recipes across 10 languages, contributed by rural and tribal communities throughout India using a specially designed mobile application. The dataset captures authentic, underrepresented culinary knowledge from digital workers—primarily rural women aged 15–45 who are native speakers of endangered languages. Each participant contributed 2–5 traditional recipes via 
a dedicated app, earning INR 750 per recipe while actively preserving their cultural heritage.

The data collection process involved recruiting local coordinators for each language, who mobilized 30–50 participants per region. Using the 
dedicated mobile application, contributors documented their traditional recipes in a structured, multimodal format that accommodated varying literacy levels through minimal text interfaces, audiovisual cues, and offline capabilities for areas with limited connectivity. The statistics of the dataset is provided in Table \ref{tab:regional_dataset}.

\begin{table*}
  \caption{Languages, States, and Dataset Statistics for Regional Recipes in India}
  \label{tab:regional_dataset}
  \begin{tabular}{c c c c c c c}
    \toprule
    \textbf{No.} & \textbf{Language} & \textbf{State (Districts)} & \textbf{Total Recipes} & \textbf{Unique Ingredients} & \textbf{Images} & \textbf{Audio Data (hh:mm:ss)} \\
    \midrule
    1 & Mundari & Jharkhand & 82 & 85 & 703 & 3:43:52 \\
    2 & Sadri & Jharkhand & 107 & 104 & 1103 & 2:32:19 \\
    3 & Santhali & Bihar & 120 & 98 & 1004 & 3:21:56 \\
    4 & Khortha & Bihar & 126 & 73 & 1129 & 4:05:33 \\
    5 & Ho & Jharkhand & 91 & 80 & 875 & 1:52:06 \\
    6 & Assamese & Assam & 113 & 148 & 1415 & 1:57:14 \\
    7 & Bodo & Assam & 95 & 190 & 1532 & 9:27:56 \\
    8 & Meitei & Manipur & 100 & 97 & 580 & 4:40:08 \\
    9 & Khasi & Meghalaya & 98 & 89 & 1928 & 6:47:40 \\
    10 & Kaman Mishmi & Arunachal Pradesh & 128 & 92 & 1129 & 7:33:27 \\
    \midrule
    -- & \textbf{Total} & --- & \textbf{1060} & \textbf{1056} & \textbf{11,398} & \textbf{46:02:11} \\
    \bottomrule
  \end{tabular}
\end{table*}

\subsubsection{Dataset Statistics - }
\label{subsubsec:dataset_Statistics}

\begin{itemize}
    \item \textbf{Total Recipes}: \textasciitilde 1,000 traditional recipes
    \item \textbf{Languages Covered}: 10 endangered languages
    \item \textbf{Geographic Coverage}: Jharkhand, Bihar, Assam, Manipur, Arunachal Pradesh and Meghalaya
    \item \textbf{Participants}: 338 rural women across all regions
    \item \textbf{Data Formats}: Text, audio recordings, and images for each recipe
    \item \textbf{Languages Distribution}:
        \begin{itemize}
            \item \textbf{Jharkhand/Bihar}: Ho, Khortha, Sadri, Santhali, Mundari
            \item \textbf{Northeast India}: Assamese, Meitei, Miju Mishmi, Bodo, Kaman Mishmi
        \end{itemize}
\end{itemize}

\subsubsection{Data Structure and Format - }
\label{subsubsec:data_structure}

Each recipe entry in the data set contains:
\begin{itemize}
    \item \textbf{Recipe Name}: In native language with transliteration
    \item \textbf{Ingredients List}: Local names
    \item \textbf{Step-by-step Instructions}: Text and/or audio in native language
    \item \textbf{Process Images}: Photo documentation of cooking steps
    \item \textbf{Cultural Context}: Traditional occasions, regional variations
    \item \textbf{Nutritional Information}: When available
    \item \textbf{Audio Recordings}: Native pronunciation and detailed explanations
\end{itemize}

\subsection{Sample Data}
\label{subsec:sample_data}

\begin{figure}[h]
    \centering
    \includegraphics[width=0.3\textwidth]{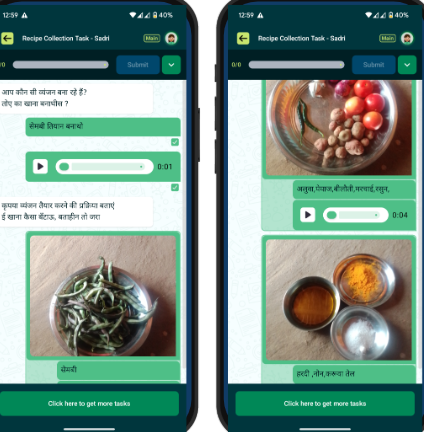}
    \caption{Data Collection App interface}
    \label{fig:app_interface}
\end{figure}

\begin{figure}[h]
    \centering
    \includegraphics[width=0.3\textwidth]{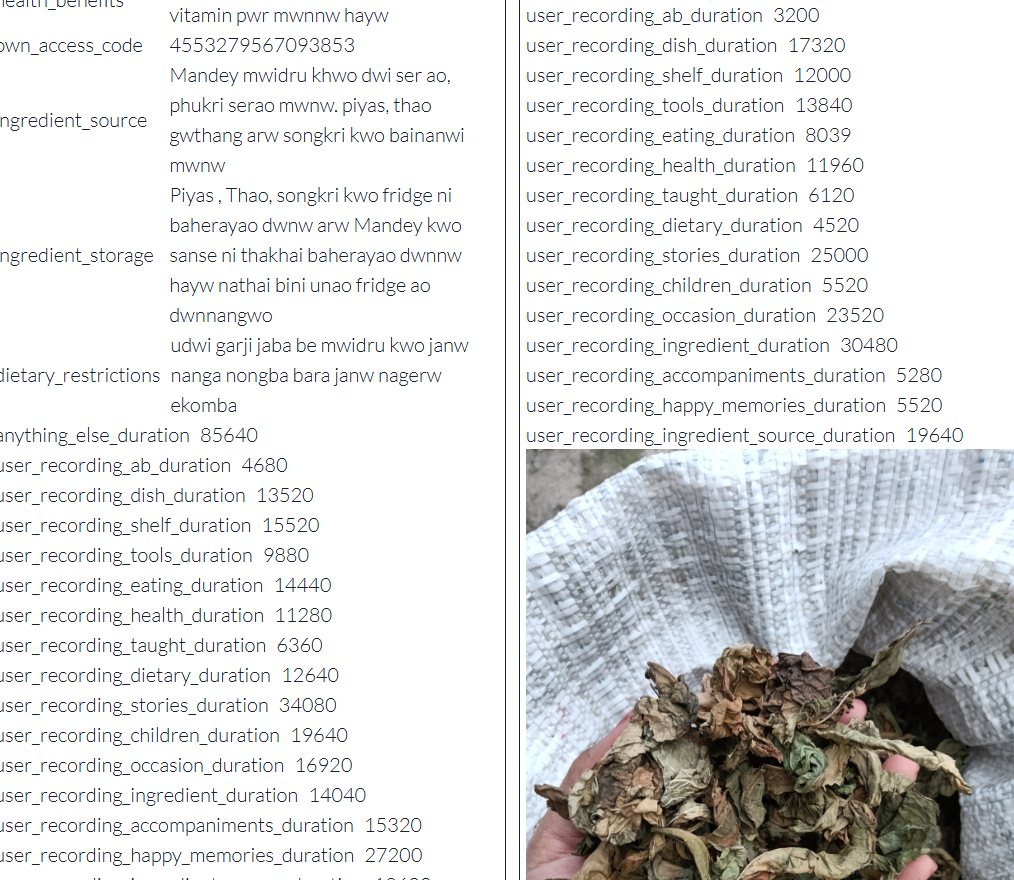}
    \caption{Raw Recipe Data in \textit{Bodo}}
    \label{fig:sample_recipe_data}
\end{figure}

\begin{figure*}[h]
  \centering
  \begin{minipage}{0.245\textwidth}
    \centering
    \includegraphics[width=\linewidth]{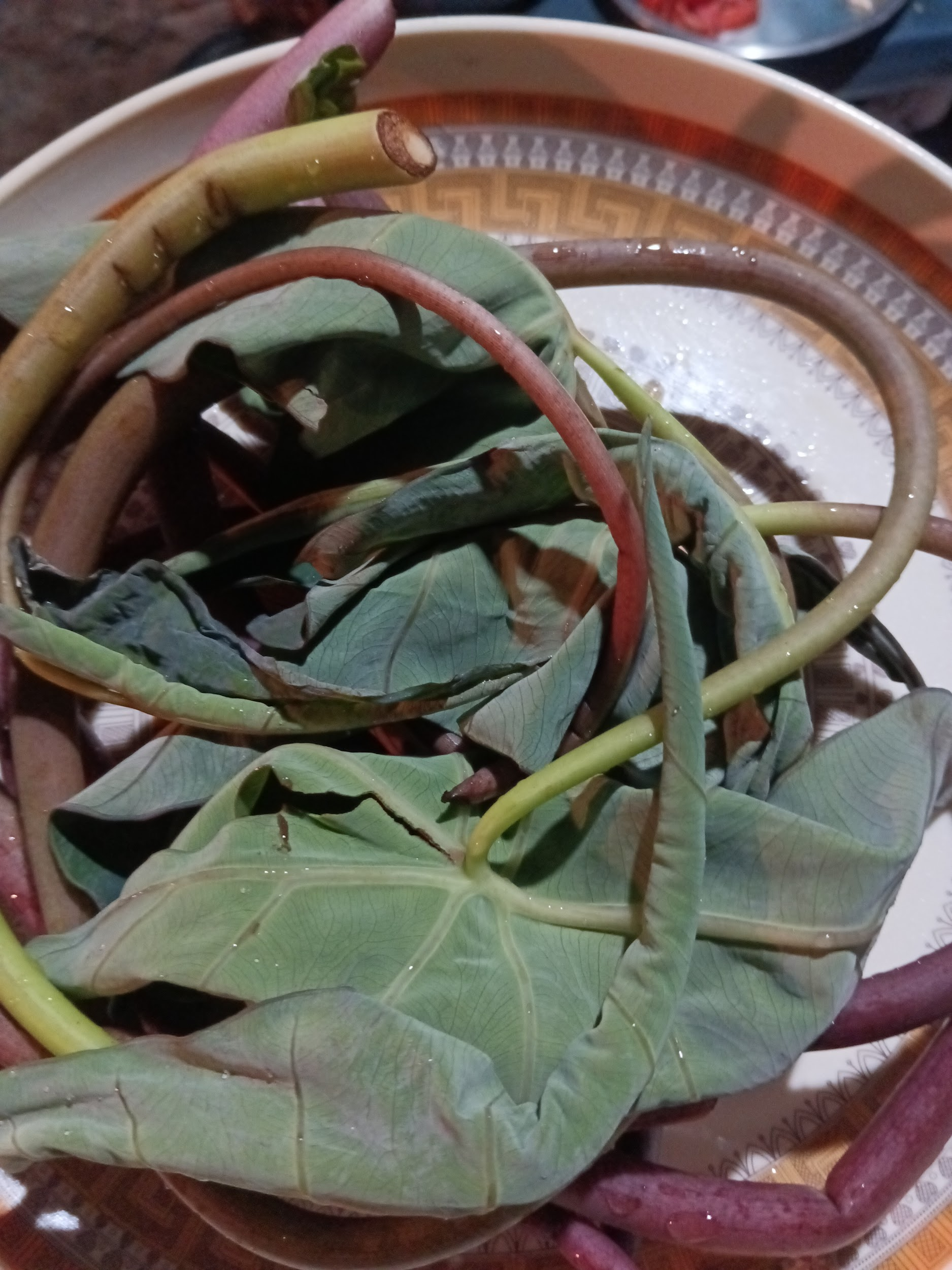}
    \textbf{(a) Raw Purple Taro (Colocasia esculenta)}
  \end{minipage}
  \hfill
  \begin{minipage}{0.245\textwidth}
    \centering
    \includegraphics[width=\linewidth]{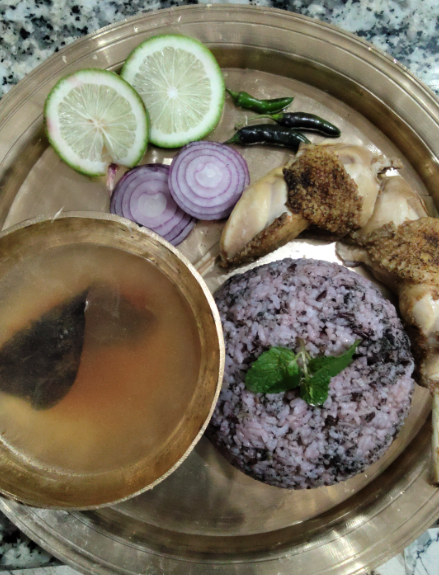}
    \textbf{(b) Boo Shat (\textit{Kaman-Mishmi} dish)}
  \end{minipage}
  \hfill
  \begin{minipage}{0.245\textwidth}
    \centering
    \includegraphics[width=\linewidth]{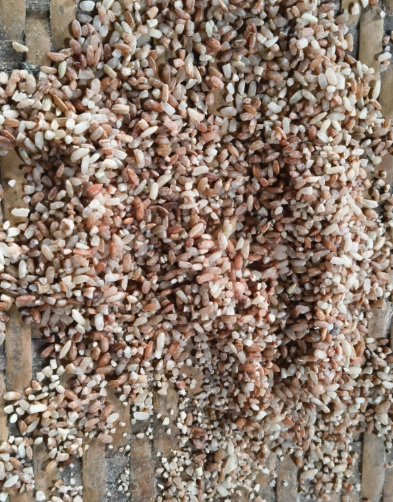}
    \textbf{(c) Red Ant Eggs (Oecophylla smaragdina)}
  \end{minipage}
  \hfill
  \begin{minipage}{0.245\textwidth}
    \centering
    \includegraphics[width=\linewidth]{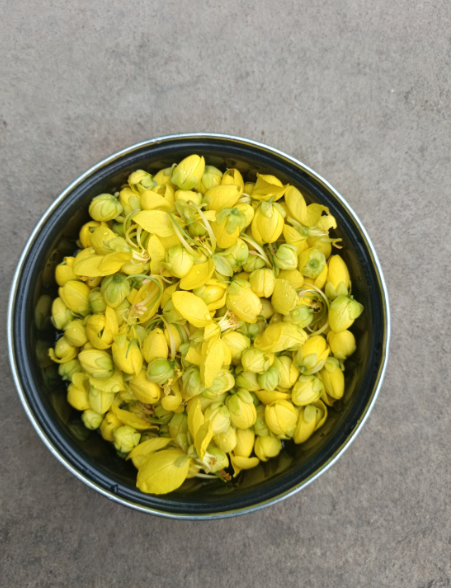}
    \textbf{(d) Amaltas (Cassia \\ fistula)}
  \end{minipage}

  \caption{Sample Images from the Indigenous Dataset}
  \label{fig:indigenous_grid}
\end{figure*}

Table \ref{tab:regional_dataset} summarizes the dataset's culinary and linguistic richness of 10 Indian communities across six states. The data highlights traditional recipes and unique ingredients, accompanied by extensive collections of images and audio recordings in each community’s native language. In total, the project documents 1,060 recipes and 11,398 unique ingredients, supported by over 46 hours of recorded audio. Figure \ref{fig:app_interface} provides a snapshot of the app interface used in the project. A sample recipe data in \textit{Bodo} is presented in Figure \ref{fig:sample_recipe_data}. Some images from the indigenous dataset are presented in Figure \ref{fig:indigenous_grid}. 

\section{Towards Building Inclusive Food Computing Infrastructures}
\label{sec:future_directions}

The multimodal community-authored recipe dataset in ten endangered Indian languages offers a rare foundation for advancing inclusive food computing. Each entry is more than a recipe—it is a culturally embedded narrative, captured through text, audio, and images, with metadata on local ingredient variants, regional adaptations, household practices, and traditional techniques.

It opens novel research directions at the intersection of AI, food culture, and low-resource language preservation, while providing structured data to inform domain-specific ontologies—e.g., concepts like \textit{“fermented millet dishes during monsoon,”} or \textit{“pregnancy-specific recipes shared through maternal oral traditions.”} Such high-context, multimodal knowledge is essential for building systems that understand food not just nutritionally or procedurally, but relationally and culturally. What follows is a discussion of how this dataset is being scaled and its potential future directions.

\subsection{Indian Foodways: Scaling Data Collection Beyond Recipes}
\label{subsec:indian_foodways}

The initial dataset of 1,000 indigenous recipes proved a successful proof of concept for real-world culinary data collection in India, but it captures only a narrow slice of the country’s vast, decentralized food knowledge systems. India’s culinary practices are deeply rooted in local agroecologies, health beliefs, ritual traditions, seasonal behaviors, and socioeconomic realities—dimensions that surface-level documentation of ingredients and cooking steps cannot adequately capture. To better capture the diversity and context of Indian foodways, we plan to expand our dataset to underdocumented indigenous regions, especially those affected by urbanization, climate change, or marginalization. The next phase will use a redesigned questionnaire with 100+ questions (up from 15) across nine modules—preparation, techniques, consumption, health, heritage, agriculture, ecology, economy, and cultural narratives. These modules will document beliefs about food-as-medicine, ingredient seasonality and sourcing, food-related livelihoods, and stories that convey emotional and intergenerational knowledge.

This framework creates a living archive of food systems that captures the cultural, ecological, and economic meanings of food beyond recipes, enabling applications in public health, sustainable agriculture, cultural preservation, and policy. The redesigned questionnaire uses emerging food ontologies to elicit machine-interpretable responses from oral traditions, ensuring consistent encoding of context, intent, timing, and health beliefs across languages and paving the way for integration with reasoning systems like FKG.in and other domain-specific knowledge graphs.

\subsection{FKG.in Integration: Enriching the Indian Food Knowledge Graph}
\label{subsec:fkg_integration}

A compelling real-world application of our culinary dataset is its potential integration with infrastructures such as FKG.in \cite{Gupta2024FKG}, India’s first dedicated food knowledge graph. 
The knowledge graph format — particularly as implemented in FKG.in — is well suited to capture the multifaceted, relational knowledge emerging from this dataset, whether in its current form or as it scales.

Unlike flat recipe databases or “universal” nutrition charts, knowledge graphs can accommodate structured and semi-structured data across domains, enabling connections between ingredients, regional names, ecological sourcing, seasonal patterns, cultural meanings, and health perceptions. Contextual data such as \textit{“prepared only during monsoon,”} \textit{“consumed after childbirth,”} or \textit{“leaf used as a plate during harvest rituals”} cannot be adequately represented without a semantic framework. Integration with FKG.in ensures usability, interoperability, and long-term impact across food computing, nutrition, sustainability, and cultural AI.

So far, FKG.in has focused on building a foundational structure linking ingredients, processes, nutrition, and dish names. However, its capacity for reasoning, multilingual querying, and inferencing across cultural, health or agroecological axes has been limited by the absence of deeply contextual, field-collected data — precisely the gap this dataset fills. It enables FKG.in to encode not only \textit{what} is cooked, but \textit{why}, \textit{by whom}, \textit{when}, and under \textit{what conditions}. This unlocks applications such as personalized health interventions rooted in traditional knowledge; culturally sensitive food recommendations; climate-aware dietary planning; and AI-driven storytelling grounded in culinary heritage. 

For example, a culturally aware cooking assistant enabled by FKG.in could infer that a fermented rice recipe shared by a \textit{Meitei} woman is not only nutritionally probiotic, but also tied to postpartum customs and seasonal harvests. While our dataset and FKG.in are separate initiatives by distinct organizations, their convergence represents a major advance in real-world, AI-driven food systems modeling — where culturally rich, ecologically grounded, and socially meaningful data finds its most powerful expression through knowledge graph integration.

\subsection{Broader Applications: Emerging Opportunities in AI, Food, and Culture}
\label{subsec:broader_applications}

Beyond integration with infrastructures like FKG.in, the dataset unlocks applications across food computing, language technologies, cultural AI, and health-oriented reasoning. Key emerging directions include:

\begin{itemize}
    \item \textbf{Procedural Soundness Checks with LLMs (Large Language Models)}: LLMs can act as evaluators, validating recipe structure by representing them as action graphs and detecting missing steps, sequencing errors, or implicit dependencies. This improves data quality, ensures completeness and assesses ease of reproduction for novice users \cite{nishimura-etal-2020-visual, yamakata-etal-2020-english, diallo2024pizza}.

    \item \textbf{Cultural Machine Translation for Endangered Languages}: The corpus described in Section \ref{subsec:recipe_project} enables training culturally grounded, domain-specific machine translation models for Indian food recipes. These models can be designed to capture procedural nuance and rich culinary context, enabling accurate, culturally sensitive translation across endangered languages \cite{yao2024cultural}.  

    \item \textbf{Cultural Recipe Corpora for Benchmarking and Supervised Fine-Tuning (SFT)}: The multimodal corpus of underrepresented Indian recipes can power culturally rich knowledge bases, diverse evaluation datasets for benchmarking models (e.g., QA, reasoning tasks, guessing games like MMLU \cite{hendrycks2020measuring} or BoolQ \cite{clark2019boolq}), and SFT for domain-specific, multimodal tasks. Such models could serve as culturally aware virtual assistants or educational tools for culinary learning, preservation, and cross-cultural exchange \cite{romero2024cvqa, winata2025worldcuisines}.

    \item \textbf{Multimodal Understanding and Generation}: Combining text, audio, and images enables applications such as generating step-by-step instructional visuals or videos from recipe text, completing cooking videos from an image or recipe prompt, recognizing indigenous ingredients from images, identifying dishes from images, or inferring intermediate steps from visual or textual cues \cite{li2024foodieqa, huang2024foodpuzzle, nishimura-etal-2020-visual}.

    \item \textbf{Reasoning-Driven Cooking Assistants}: Culturally \\grounded data supports contextual reasoning assistants that could explain \textit{why} a step matters, \textit{what} happens if altered, or \textit{how} to adapt recipes to allergen or ingredient availability constraints - combining common sense reasoning, cultural sensitivity, and procedural awareness. These systems can help users navigate real-world cooking constraints while supporting the creative adaptation and evolution of traditional recipes across diverse and global culinary contexts \cite{Shirai2021, hu-etal-2024-bridging}.
\end{itemize}

\section{Challenges and Open Questions}
\label{sec:challenges}

Some valuable learnings from this initiative, which may guide similar future efforts, are outlined below: 

\begin{itemize}
    \item \textbf{Trust-building}: Initial skepticism stemmed from disbelief that digital work—particularly involving recipes and storytelling—could result in real payment. Many participants were unfamiliar with such tasks being seen as valuable or remunerative, necessitating reassurance and clear communication through trusted local coordinators. 
    \item \textbf{Seasonality constraints}:Many tribal and rural recipes rely on foraged or locally grown ingredients that are only available during specific seasons. As a result, a significant number of off-season recipes and ingredients could not be captured in the initial dataset.
    \item \textbf{Infrastructure limitations}: Unreliable internet connectivity and power supply in remote villages posed challenges for uploading multimodal data. To mitigate this, the mobile application was designed to support offline recording, requiring internet access only during final submission.
    \item \textbf{Payment equity}: Recipe complexity varied dramatically — from elaborate dishes involving early-morning foraging and hours of ancestral techniques, to simpler but equally authentic preparations. Determining fair compensation while honoring the cultural value of all contributions proved to be a nuanced challenge.
    \item \textbf{Data quality}: Ensuring consistency across diverse languages, cooking styles, and literacy levels required a two-layer validation process involving local coordinators and project managers. This system helped ensure data completeness and cross-regional coherence, ultimately strengthening the participatory framework and underscoring the importance of patient, hyper-local approaches to digital inclusion.
\end{itemize}

\section{Conclusion}

 The present work describes a dataset of 1,000 indigenous recipes, collected in ten languages in Eastern and Northeast India. 
 Data collection and remuneration were facilitated through a smartphone-based data collection platform. The collection process involved unprecedented outreach efforts, including the participation of language experts, community coordinators, and data scientists. The primary data contributors were women who are native speakers of these ten languages.

Although originally designed for cultural inclusion and creation of knowledge bases, the dataset serves as a valuable resource for food computing by presenting an exhaustive list of cooking practices - such as rare ingredients, seasonal traditions, rituals, and nutritional content - unique to home cooking and often overlooked in industrially structured recipe formats. The data set also holds potential for developing culturally aware large-language models (LLMs), while the English translations can serve as parallel corpora for domain-specific machine translation. Future work will involve expanding the dataset to include additional rare and under-resourced languages of India.

\begin{acks}

 We gratefully acknowledge the local coordinators and contributors whose time, knowledge, and effort made this work possible. We are thankful to the Mphasis AI \& Applied Tech Lab at Ashoka - a collaboration between Ashoka University and Mphasis Limited - for their support.
\end{acks}

\bibliographystyle{ACM-Reference-Format}
\bibliography{main}










\end{document}